# Tetragonality of Fe-C martensite – a pattern matching electron backscatter diffraction analysis compared to X-ray diffraction

Tomohito Tanaka[a,b,*], Naoki Maruyama[b], Nozomu Nakamura[c], Angus J. Wilkinson[a]

[a] Department of Materials, University of Oxford, Parks Road, Oxford, UK
[b] Nippon Steel Corporation, 20-1 Shintomi, Futtsu, Chiba, Japan
[c] NS Solutions Corporation, 20-1 Shintomi, Futtsu, Chiba, Japan

* corresponding author.
E-mail: tanaka.m9p.tomohito@jp.nipponsteel.com

Abstract

Measurements of the local tetragonality in Fe-C martensite at microstructural length-scale through pattern matching of electron backscatter diffraction patterns (EBSPs) and careful calibration of detector geometry are presented. It is found that the local tetragonality varies within the complex microstructure by several per cent at largest and that the scatter in the axial ratio is increased at higher nominal carbon content. At some analysis points the local crystal structure can be regarded as lower symmetry than simple body centred tetragonal. A linear relation between the nominal carbon content and averaged local tetragonality measured by EBSD is also obtained, although the averaged axial ratio is slightly below that obtained from more classical X-ray diffraction measurements.

## 1. Introduction

An increasing demand for the development of higher strength steel drives us to gain fundamental knowledge on the microstructure and mechanical properties of Fe-C martensitic steels. The crystal structure of Fe-C quenched martensite analysed by X-ray diffraction (XRD) is either body centred cubic (BCC) or body centred tetragonal (BCT) and the tetragonality, *c/a*, has been found to be linearly dependent on the carbon content ([C] in wt.%) as follows [1-6]:

$$c/a = 1 + 0.045 \, [C] \tag{1}$$

However, despite intensive studies for nearly 90 years since the first discovery of the tetragonality in Fe-C martensite [7, 8], there is still ongoing debate on several issues regarding the minimum carbon content that gives rise to tetragonality [9-16], factors affecting abnormally high/low tetragonality seen in some Fe-C alloys [17-26], and spatial variation in tetragonality [23, 27-30]. For instance, Sherby *et al* mentioned that the tetragonality starts to appear when [C] exceeds 0.6 wt.%, which was ascribed to the solubility of carbon in hexagonal martensite [10, 11]. The hexagonal martensite was assumed to be formed as an intermediate phase during face centred cubic (FCC) to BCT transformation, although the presence of the intermediate phase has not been confirmed by experiment. Hutchinson *et al* also reported that the tetragonality is not observed for 0.1~0.5 wt.% C quenched martensite [12]. Xiao *et al* reported an abrupt change in the tetragonality at ~0.55 wt.% C while the transition BCC↔BCT crystal structure itself occurs at 0.18 wt.% C [13].

The difficulty in analysing the tetragonality of low carbon martensite stems from carbon redistribution during cooling (*i.e.* auto-tempering) leading to a peak overlap in the XRD profile for α-Fe. To suppress auto-tempering, Cadeville *et al* used a splat cooling [9] and the *c/a* expected from Eq. (1) was obtained even in Fe-0.2 wt.% C martensite. Recently Lu *et al* applied the Rietvelt refinement analysis [15] in order to deconvolute the α-Fe {200} XRD peaks concluding that the slope in the Eq. (1) is rather smaller (0.031 instead of 0.045) and the *c/a* > 1 was obtained even in 0.12 wt.% C martensite, although the detail in the Rietvelt refinement was not mentioned in the literature [15]. Therefore, the experimentally determined minimum carbon content that gives rise to the tetragonality appears to be dependent on the cooling rate from austenitising temperature and the sophistication of the XRD line profile analysis method employed.

In the XRD profile, the evidence of tetragonality can be seen in the {200} diffraction peak which starts to split into two visible peaks ((200)/(020) and (002)) when the carbon content is beyond the threshold. Some authors reported the presence of the third peak between (200)/(020) and (002) two main peaks [23, 27], suggesting that the third peak is derived from lower tetragonal or 'local' orthorhombic martensite in the bulk BCT martensite [23, 27]. It is also reported that the area of the third peak is increased during room temperature (RT) aging [27]. Chen *et al* [23] speculated that the local perturbation in the tetragonality might be derived from the large residual stresses present at plate intersections as the third peak is very evident when the martensite is formed from polycrystalline austenite while it is not clearly seen when the martensite is formed from single crystalline austenite. Zener [14], Maugis [31], and Chirkov *et al* [32] also mention that such local residual stresses can alter carbon occupation (*i.e.* carbon content in one of the sublattices of the octahedral interstitial site, namely Zener ordering [14]), which would decrease local tetragonality. Furthermore, recent atom probe tomography (APT) analysis revealed the heterogeneous distribution of carbon concentration even in as-quenched Fe-C martensite due to the relatively high *Ms* temperature (martensite transformation start temperature) for low carbon martensite [12, 33, 34]. Morsdorf *et al* revealed the extent of auto-tempering is different according to the size of laths [33]. Then it is natural to expect that the tetragonality should be varied within the complex and hierarchical microstructure in Fe-C martensitic steels. However, the direct observation of spatial variation in tetragonality by means of laboratory XRD apparatus is not possible due to the lack in the spatial resolution.

Electron backscatter diffraction (EBSD) complements XRD and is capable of analysing crystal structure at a few tens of nanometre scale and has been widely used for characterising the crystallographic features of Fe-C martensite, such as block/packet size [33, 35-39], variant selection [40-42] and prior austenite crystal orientation [43-47]. However, the lack in angular resolution of conventional EBSD analysis and poor accuracy in calibrating camera geometry hinder the differentiation of BCT structure from BCC. It is common in this research field to analyse the crystal orientation of martensite through conventional EBSD indexing assuming that the crystal structure is BCC (BCC indexing) [48, 49]. Stormvinter [48] tried to index crystal orientation of Fe-C quenched martensite with several carbon contents by conventional EBSD analysis using Hough transform assuming that the crystal structure is BCT (BCT indexing) with fixed tetragonality. It is found that the BCT indexing success rate of nearly 100% is achieved only when [C] is 1.80 wt.% with expected *c/a* ~ 1.08 while it is decreased as the carbon content decreases. When the carbon content is 0.75 wt.%, the success rate becomes < 50 % meaning that *c*-axis cannot be distinguished reliably from *a, b*-axes [48]. In order to characterise the tetragonality and its spatial variation in Fe-C martensite, therefore, it is necessary to improve angular resolution of EBSD analysis.

High angular resolution EBSD (HR-EBSD) analysis is now possible with angular resolution of 0.006 degree (strain sensitivity $\sim 10^{-4}$ or better) using cross-correlation analysis of EBSD patterns (EBSPs) [50, 51]. However, HR-EBSD measures *relative* strain compared to the strain at a reference point which must be known for *absolute* strain analysis. In general, the strain state at the reference point is not known *a priori* resulting in the so-called reference pattern problem which has limited HR-EBSD to relative strain mapping. Recently, the authors used dynamically simulated EBSPs as a true reference pattern for *absolute* strain analysis with careful determination of a pattern centre (PC) position [52], and it was successful in determining absolute strain near an indent in Fe irrespective of the location of reference points with strain measurement accuracy of the order of $10^{-4}$. This level of accuracy is certainly enough to analyse the tetragonal distortion of martensite of the order of $\sim 10^{-2}$. However, the calibration method used in [52] is likely to give the best accuracy in PC determination when the experimental pattern is obtained from an unstrained area as deformation gradient tensor is not taken into account for the dynamical simulation.

Here we propose a route to determining PC position through the pattern matching even in the case where strain free area is not available such as in Fe-C martensite. Once the PC position is determined it is possible to simulate EBSD patterns for BCT Fe with changing *c/a,* thereby allowing for the optimisation of *c/a* by searching the best fit simulated pattern to experimental counterpart [29, 53]. To verify our EBSD tetragonality analysis results, XRD analysis was also performed. The method was used to determine the local tetragonality in a series of Fe-C martensite samples with nominal carbon content varying from 0.07 wt% to 1.29 wt%.

## 2. Methodology

2-1. Materials

The chemical composition of Fe-C alloys prepared in this study is listed in Table 1. All alloys were vacuum melted, hot rolled and cold rolled down to the thickness of approximately 1 mm. Then all Fe-C alloys except for the interstitial free (IF) steel were austenitised and water-cooled to obtain martensite microstructure. The austenitisation temperature is also listed in Table 1. Sub-zero treatment using liquid nitrogen was followed to reduce the amount of retained austenite. After that, the martensite specimens were kept at room temperature until XRD and EBSD measurements for 0.5 ~ 2 years. After the cold rolling process, the IF steel was annealed at 800 °C for 10 minutes for recrystallisation. This material was prepared to obtain standard XRD and EBSD patterns from BCC ferrite.

All specimens were cut and polished with SiC papers followed by colloidal silica (Bheuler, MasterMet) polishing. The microstructure for each material was observed with a Scanning Electron Microscope (SEM, Zeiss Merlin). Fig. 1 shows the microstructure and the observed morphology of martensite is listed in Table 1. After the SEM observation, the IF steel was annealed at 700 °C for 30 minutes in vacuum to remove potential residual strain introduced by the polishing.

Table 1. The materials used in this study. Ms temperature is calculated using the equation listed in [54].

| Alloy composition [wt.%] | Austenitising temperature [°C] | Ms [°C] | Martensite morphology | RT aging time [month] |
|---|---|---|---|---|
| IF steel (0.0049C) | - | - | (ferrite) | - |
| 0.07C-1.0Mn | 1100 | 444 | lath | 3 |
| 0.24C-1.0Mn | | 390 | lath | 3 |
| 0.44C-1.0Mn | | 326 | lath | 6 |
| 0.59C-1.0Mn | | 279 | lath, plate | 4 |
| 0.77C-1.0Mn | | 222 | lath, plate | 3 |
| 1.00C | 910 | 182 | lath, plate | 24 |
| 1.29C | | 90 | plate | 23 |

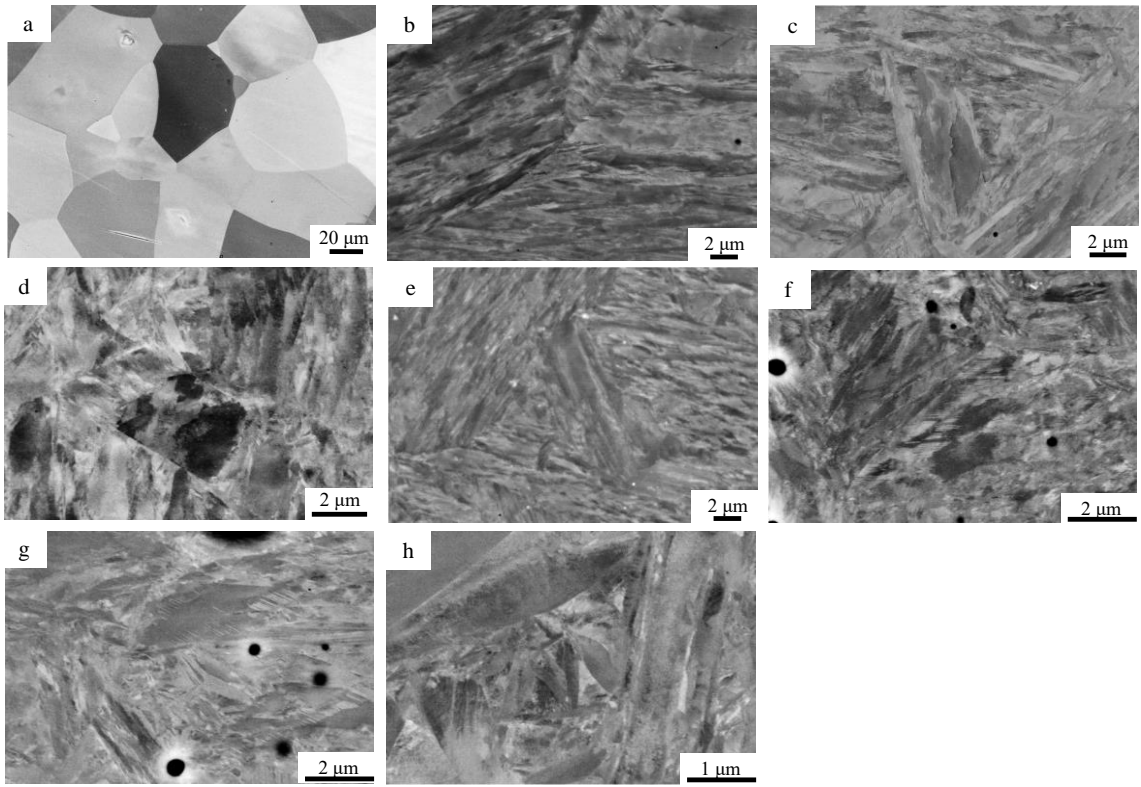

Figure 1. Electron channelling contrast images for (a) IF steel, (b) Fe-0.07 wt.% C, (c) Fe-0.24 wt.% C, (d) Fe-0.44 wt.% C, (e) Fe-0.59 wt.% C, (f) Fe-0.77 wt.% C, (g) Fe-1.00 wt.% C, and (h) Fe-1.29 wt.% C martensitic steels.

2-2. XRD measurements

XRD measurements were performed at RT using a Rigaku SmartLab diffractometer equipped with a Co tube. The voltage applied to the tube was 40 kV. The diffracted intensity was measured in the range 10 to 140 degrees for $2\theta$ with a step size of 0.02 degree $2\theta$.

The Fe {200} peaks from Fe-C martensitic steels were deconvoluted in the following manner. The background was fitted to a cubic spline and subtracted from the original profiles. A pseudo-Voigt function was used for the peak fitting procedure. Since the incident X-ray was not monochromated, a pair of two peaks derived from Co-K$\alpha_1$ and Co-K$\alpha_2$ irradiation forms one apparent peak. The area ratio of the peak excited by Co-K$\alpha_1$ to the peak excited by Co-K$\alpha_2$ irradiation was set to be 2. Two or more pairs of peaks were necessary to reproduce experimental XRD Fe {200} profiles when [C] was greater than a threshold value.

2-3. EBSD measurement

The EBSPs were collected using an SEM-EBSD system (SEM: JEOL 7100F, EBSD: EDAX DigiView camera). The sample was inclined 70° from the horizontal and the phosphor screen was tilted 3° from the vertical. The acceleration voltage of the field emission gun was 20 *kV* with beam current of 11~14 *nA*. The EBSPs were recorded at 12 bit depth and saved as tiff images with 956 × 956 pixels resolution using a circular phosphor screen. Static background and/or dynamic background were subtracted from the raw EBSP. The crystal orientations were first analysed, assuming a BCC crystallography, using conventional Hough transform based analysis with *TSL OIM Data Collection* software (EDAX) or *DynamicS* (Bruker) software and later using the pattern matching of EBSPs [52].

2-3-1. Calibration of camera geometry for the pattern matching of EBSPs

Fig. 2 shows the geometrical configuration between crystal and an EBSP imaging screen. Suppose that the [001] *c*-axis of BCC Fe is perpendicular to the image screen and Point A ($a_1$, $a_2$, $a_3$) on the (001) plane is projected at Point P, $DD \times (a_1/a_3, a_2/a_3, 1)$, on the screen, where *DD* denotes a camera length. When the *c*-axis is elongated by a factor of *t*, Point A moves to Point A'($a_1$, $a_2$, $ta_3$). The resultant projection coordinate of Point P' on the screen is $DD/t \times (a_1/a_3, a_2/a_3, t)$. In this particular case a decrease in camera length or an increase in the tetragonality can make the shift in projected point invariant, meaning that the uncertainty in determining a camera length (usually calibration precision ~ 0.5 % of a screen width [55]) is directly linked to tetragonality measurement errors.

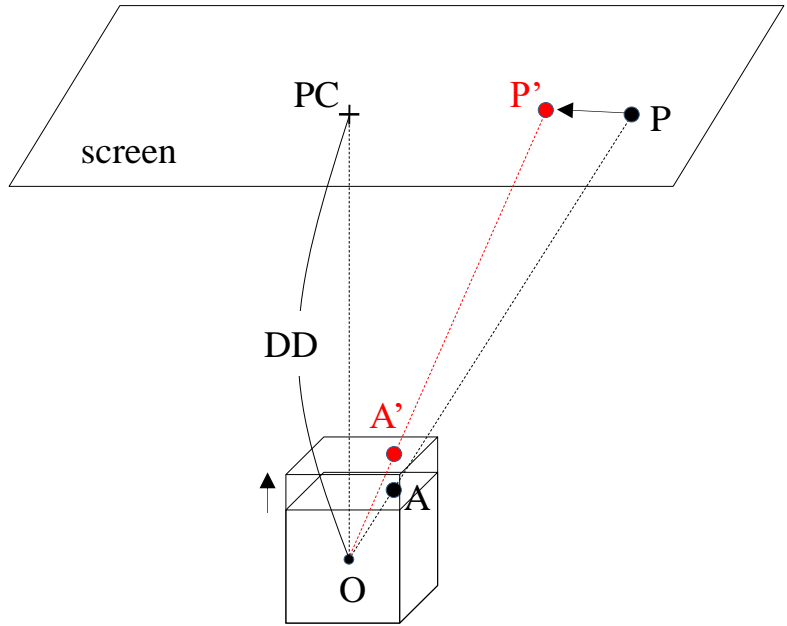

Figure 2. Geometrical configuration between crystal and an imaging screen. Point O is the point in which the diffracted electrons are generated.

The authors developed a pattern matching approach for the accurate and precise calibration of EBSD system using a global optimisation algorithm [52]. This approach requires experimental EBSPs obtained from an unstrained area for the best accuracy in the calibration. Since strain-free area is not known *a priori* in Fe-C martensite, a thin foil of a strain-free specimen with dimensions of 20 *μm* × 20 *μm* × 1 *μm* (thickness: 1 *μm*) was mounted onto the martensite specimen (Fig. 3(a)). The thin foil was extracted from well-annealed IF steel using a Ga focused ion beam (FIB) micro-sampling apparatus (Hitachi FB2000). The acceleration voltage of Ga ion gun was 40 *kV*, followed by 10 *kV* to reduce the thickness of Ga ion beam damaged layer. The extracted thin foil specimen was glued on the martensite surface by depositing Pt or W layer at the edge of the foil (Fig. 3(a)). After that, in order to remove Ga-ion beam damaged layer near the foil surface, broad Ar ion beam with acceleration voltage of 1 *kV* was employed for 5 minutes. Figs.3(b)-3(d) shows how the Ar ion beam sputtering improved the quality of EBSP from the thin foil. As the Ar beam irradiation time increases an improvement in the pattern quality becomes very evident.

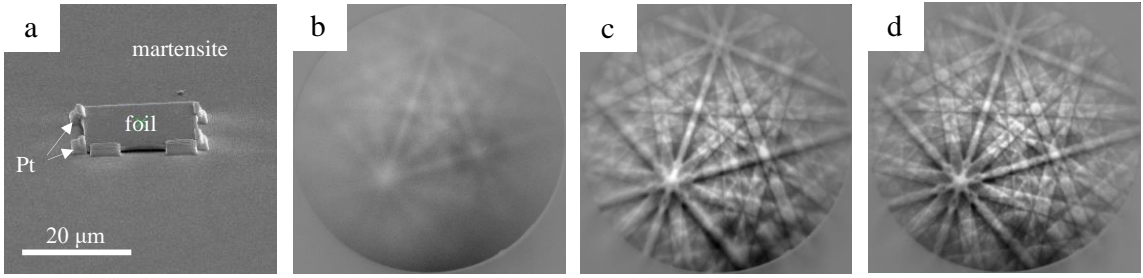

Figure 3. (a) A foil extracted from annealed IF steel placed onto martensite specimen. (b)-(d) EBSPs obtained from the foil surface after 1 *kV* Ar ion beam irradiation of (b) 0 *s*, (c) 100 *s*, (d) 300 *s*.

Such a good quality pattern can be used for the calibration through pattern matching analysis. The experimental pattern is compared to dynamically simulated EBSPs for BCC Fe (lattice parameter = 0.2866 *nm*) with changing PC positions and crystal orientations (*i.e.* Euler angles, ($\varphi_1$, $\Phi$, $\varphi_2$)). The PC coordinate ($PC_x$, $PC_y$, $PC_z$) is determined by finding the best matched simulated pattern determined by maximising the normalised cross-correlation between the experimental and simulated patterns.

When the electron beam is impinged on the martensite surface by shifting a beam position from the thin foil, the PC position is changed accordingly as follows (Fig. 4(a)):

$$PC_x^M = PC_x^F + \frac{\Delta x}{\eta W} \quad (2a)$$

$$PC_y^M = PC_y^F + \frac{\Delta y \cos\tau}{\eta H} \quad (2b)$$

$$PC_z^M = PC_z^F - \frac{1}{\eta H}\left(\Delta y \sin\tau - \frac{s}{\cos\tau}\right) \quad (2c)$$

where $PC^M$, $PC^F$ are the PC positions when analysing the martensite and the foil, respectively. $\eta$ is the pixel size and *H*, *W* correspond to the number of pixels along the pattern height and width. $\Delta x$ and $\Delta y$ are the beam shift across the sample surface from the analysis point on the foil. $\tau$ is the tilt angle between the sample and the screen ($\tau$=90-$\tau_{sample}$+$\tau_{detector}$, where $\tau_{sample}$ is the sample tilt, 70 degrees, and $\tau_{detector}$ is the detector tilt, 3 degrees in this study). *s* is the thickness of the foil.

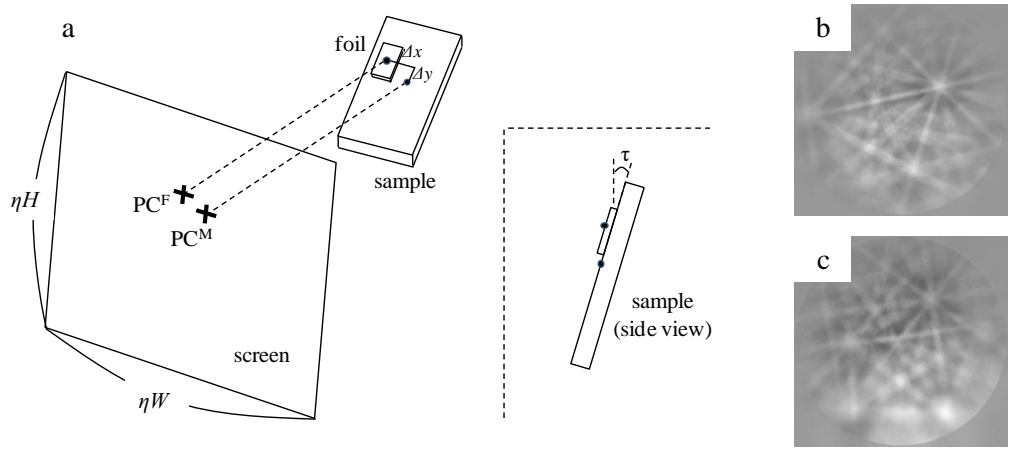

Figure 4. (a) Geometrical relationship between an image screen and sample/foil position. (b) A typical EBSP from Fe-C martenstite. (c) An example of overlapped pattern.

It is noted here that electron beam shift should be used to change an analysis point instead of sample stage shift since the stage movement is not always perfect [55].The electron beam shift should be within a certain range, say a few hundred μm so as to avoid the distortion in the beam shift [56].

In this way care was taken to assess the PC position for accurate and precise determination of *c/a* ratio. More than 10 EBSPs for each martensite sample were analysed and the location of the analysis points is shown in the Supplementary section. A typical EBSP from Fe-C martensite is shown in Fig. 4(b). The analysis points were randomly selected and separated from each other by more than a few *μm*, and shown in Supplementary materials. Therefore, all the EBSPs should be obtained from different blocks. Some patterns were non-indexable, very blurred or overlapped (such as that shown Fig. 4(c)) but these were rejected and not used for the pattern matching analysis.

2-3-2. Pattern matching of EBSD patterns from Fe-C martensite for tetragonality measurement

With the fixed camera geometry determined in the previous section, a similar pattern matching was performed between EBSPs from the martensitic steels and dynamically simulated reference patterns for BCT Fe with different *c/a* ratios. Since the PC position is fixed, the number of variables to be optimised is four ($\varphi_1$, $\Phi$, $\varphi_2$, *c/a*). An initial guess of the Euler angles is given randomly within the prescribed range between the Euler angles determined by Hough based conventional analysis ± 5°. Since Hough transform based analysis may fail to distinguish the *c*-axis from the *a*- and *b*-axes [48], the following matrix $S_{axis}$ is multiplied by the rotation matrix, *R*, in turn in order to consider the possibility of *c*-axis being along each of the three principal axes.

$$R = \begin{pmatrix} \cos\varphi_2 & \sin\varphi_2 & 0 \\ -\sin\varphi_2 & \cos\varphi_2 & 0 \\ 0 & 0 & 1 \end{pmatrix} \begin{pmatrix} 1 & 0 & 0 \\ 0 & \cos\Phi & \sin\Phi \\ 0 & -\sin\Phi & \cos\Phi \end{pmatrix} \begin{pmatrix} \cos\varphi_1 & \sin\varphi_1 & 0 \\ -\sin\varphi_1 & \cos\varphi_1 & 0 \\ 0 & 0 & 1 \end{pmatrix} \quad (3a)$$

$$S_{axis} = \begin{pmatrix} 1 & 0 & 0 \\ 0 & 1 & 0 \\ 0 & 0 & 1 \end{pmatrix}, \begin{pmatrix} 0 & 0 & 1 \\ 0 & 1 & 0 \\ -1 & 0 & 0 \end{pmatrix}, \begin{pmatrix} 1 & 0 & 0 \\ 0 & 0 & 1 \\ 0 & -1 & 0 \end{pmatrix} \quad (3b)$$

where the first matrix of $S_{axis}$ is identity matrix, the second matrix swaps *a*-axis with *c*-axis and the third matrix swaps *b*-axis with *c*-axis while maintaining the right-handed co-ordinate. Then $S_{axis} R$ is used to extract a gnomonic projection image from the master patterns for BCT Fe with different *c/a*, which were simulated by *DynamicS* software (Bruker) [57] with lattice parameter of *a,b*-axes being 0.2866 *nm* and of *c*-axis ranging from 0.2826 to 0.3056 *nm*. The similarity between experimental EBSPs from the martensite and simulated patterns was assessed using a normalised cross correlation coefficient, *r*. Then, for each *c/a* ratio simulated, the given Euler angles are refined through the optimisation process using a differential evolution algorithm [52, 58] so that the maximum of *r* is obtained.

## 3. Results and Discussion

3-1. XRD study on the tetragonality of Fe-C quenched and RT aged martensite

The XRD profiles for IF steel and Fe-C quenched- RT aged marntensite with different [C] are shown in Fig. 5(a). The sign of tetragonality can be seen in $\alpha$-Fe XRD peaks (Figs. 5(b)-(e)) when the carbon content is high enough as a form of the peak splitting. Then the detailed structure of $\alpha$-Fe {200} diffraction profile is analysed as the peak from $\gamma$-Fe is not overlapped. In order to determine the minimum carbon content that gives rise to tetragonaltiy ($c/a > 1$), {200} diffraction peak asymmetry factor, $A_f$, was assessed as follows:

$$A_f = \frac{m}{n} \tag{4}$$

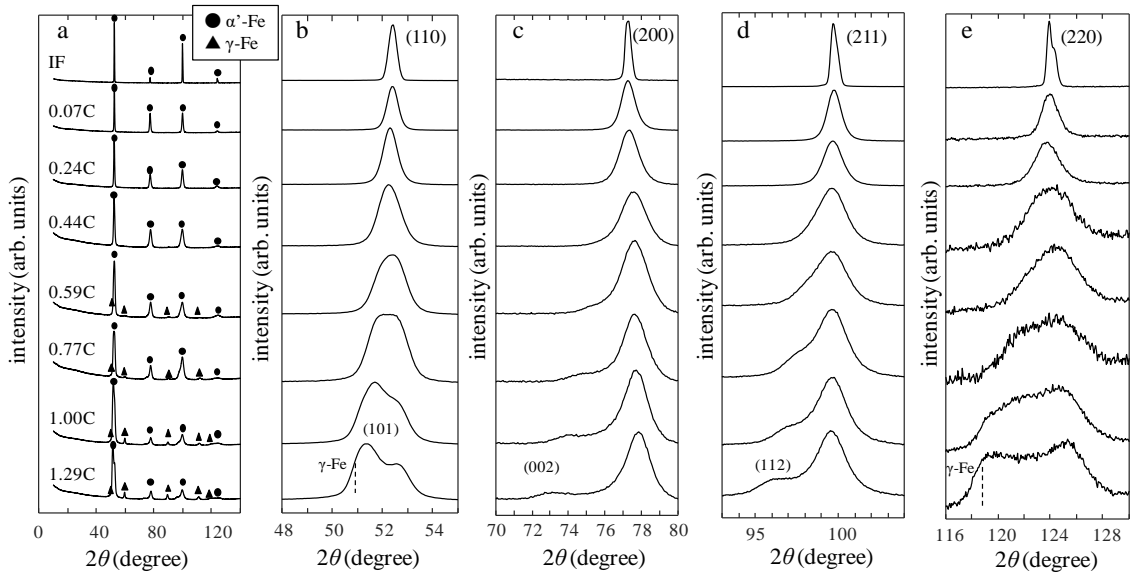

Figure 5. XRD profiles from quenched and RT aged Fe-C martensite and IF steel. (a) wide scan XRD profile. XRD profile for (b) $\alpha$-Fe {110}, (c) $\alpha$-Fe {200}, (d) $\alpha$-Fe {211}, (e) $\alpha$-Fe {220}.

where $m, n$ represents the width of the peak at 10% peak height, measured to the left, and right respectively of the peak maximum position (as shown in the inset in Fig. 6). Fig. 6 shows the asymmetry factor for $\alpha$-Fe {200} profile as a function of carbon content.

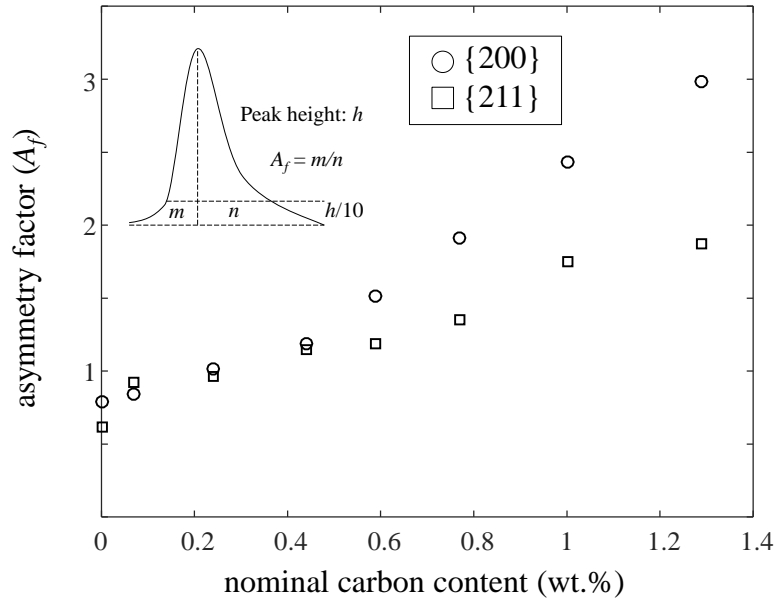

Figure 6. Asymmetry factor, $A_f$, for $\alpha$-Fe {200}, {211} XRD peak as a function of the nominal carbon content

At [C] = 0.07 wt.%, the peak profile shows a tailing nature with $A_f < 1$. This is because the XRD peak excited by Co-K$\alpha_2$ is overlapped to the right (higher scattering angle side) of the peak excited by Co-K$\alpha_1$. When [C] = 0.24 wt.%, the profile becomes symmetric because of the broadening of the peak width. From [C] = 0.44 wt.%, the peak showed a fronting nature ($A_f > 1$) indicating that the tetragonal martensite starts to appear. This feature is identical to the $\alpha$-Fe {211} profile too. It is concluded, therefore, that the minimum carbon content that gives rise to the tetragonality is between 0.24 – 0.44 [C] for the materials prepared in this study.

Fig. 7 shows the detailed profile of $\alpha$-Fe {200} in addition to the deconvoluted peaks. The fitted peak positions are listed in Table 2. Since the incident X-ray was not monochromated, a pair of two peaks derived from K$\alpha_1$ and K$\alpha_2$ irradiation forms one apparent peak. As is noted before, one apparent peak (Peak I) is enough to describe the profile from Fe-0.07, 0.24 wt.% C martensite (Figs. 5(b) and 6) as the crystal structure is BCC. When [C] is beyond 0.24 wt.%, one more peak from (002) diffracting plane (Peak II) needs to be introduced. The full width at half maximum (FWHM) for each peak is set to be identical on the assumption that the extent of residual strain acting on *a, b, c*-axes is almost the same amount. Then, as indicated by ref. [23, 27], additional peak(s) in between Peaks I&II is necessary to fully account for the experimental XRD profile from Fe-C martensite with [C] ≥ 0.44 wt.%. The peak positions and FWHM for the additional peaks can be determined in a variety of ways as the number of peaks in between Peaks I&II is uncertain. Therefore, the residual obtained by subtracting the area for Peaks I&II from original peak area is calculated as a metric to describe the deviation from 'pure' BCT with single tetragonality.

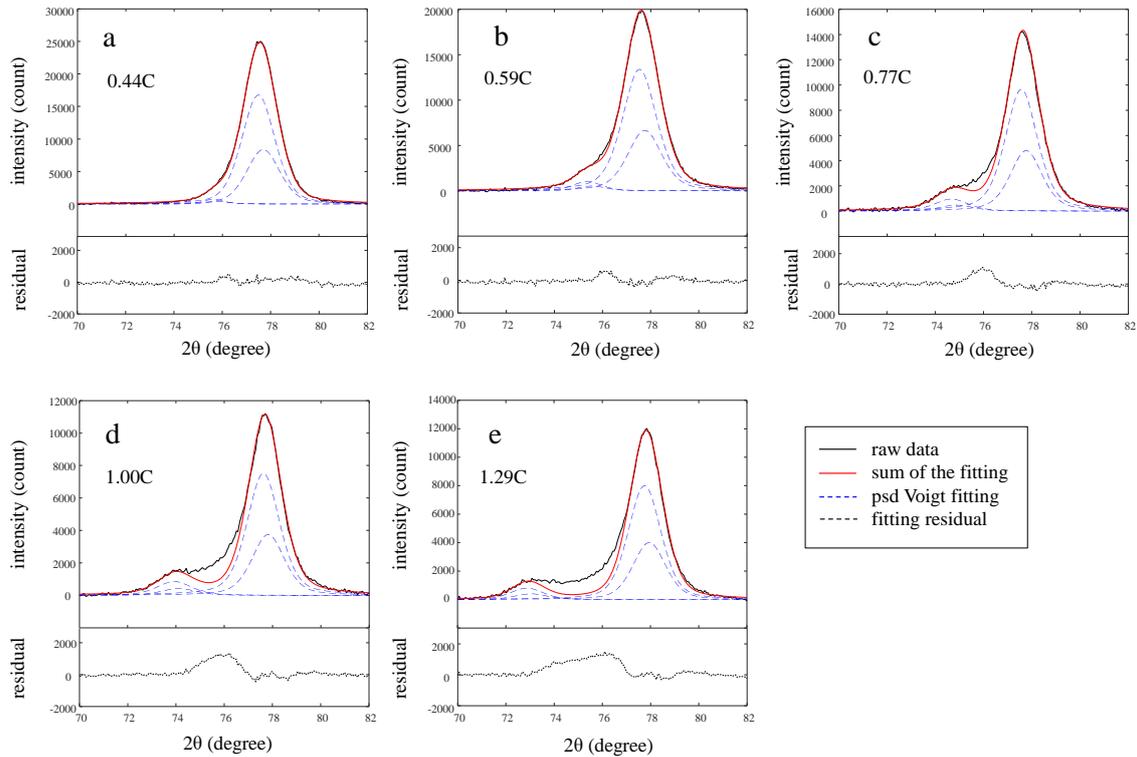

Figure 7. $\alpha$-Fe {200} XRD line profile along with the fitted results and fitting residuals for (a) Fe-0.44 wt.%C, (b) Fe-0.59 wt.%C, (c) Fe-0.77 wt.%C, (d) Fe-1.00 wt.%C, and (e) Fe-1.29 wt.%C.

The peak area ratio of Peak II to Peak I, $A_{II}/A_I$, is not 0.5 which is expected from the multiplicity ratio (Table 2). Keating and Goland [59], Moss [60] demonstrated that the displacement of Fe atoms from their mean lattice sites caused by interstitial carbon leads to a decrease in the XRD peak intensity of (002). Furthermore, the presence of the residual (*i.e.* Peak III) also contributes to a decrease in $A_{II}/A_I$. It was proposed that the occurrence of Peak III is caused by a redistribution of carbon interstitials over $O_x$, $O_y$ and $O_z$ octahedral interstices leading to a formation of less tetragonal or local orthorhombic region [23]. Fig. 8 shows the area of fitting residual as a function of the nominal carbon content. When [C] $\geq$ 0.44 wt.%, the residual is increased according to the carbon concentration, suggesting that the area with less tetragonal or local orthorhombic region is increased. Then, the 'averaged' tetragonality within an XRD analysis volume, $(c/a)_{ave}$, should be smaller than the tetragonality determined only by the Peaks I&II positions, $(c/a)_{II/I}$.

Table 2. α-Fe {200} XRD peak fitting result.

|  | IF | 0.07C | 0.24C | 0.44C |  | 0.59C |  |
|---|---|---|---|---|---|---|---|
| Peak No | I | I | I | I | II | I | II |
| 2θ (K$\alpha_1$) [deg] | 77.207 | 77.210 | 77.259 | 77.490 | 75.634 | 77.536 | 75.355 |
| Area ratio (Peak II/I) | - | - | - | - | 0.044 | - | 0.038 |
| axis length [nm] | 0.2867 (a, b, c) | 0.2867 (a, b, c) | 0.2866 (a, b, c) | 0.2858 (a, b) | 0.2918 (c) | 0.2857 (a, b) | 0.2927 (c) |
| c/a (II/I) | 1 | 1 | 1 | 1.020 | | 1.024 | |

|  | 0.77C |  | 1.00C |  | 1.29C |  |
|---|---|---|---|---|---|---|
| Peak No | I | II | I | II | I | II |
| 2θ (K$\alpha_1$) [deg] | 77.565 | 74.663 | 77.617 | 73.925 | 77.756 | 72.867 |
| Area ratio (Peak I = 1) | 1 | 0.100 | 1 | 0.113 | 1 | 0.100 |
| axis length [nm] | 0.2856 (a, b) | 0.2950 (c) | 0.2854 (a, b) | 0.2975 (c) | 0.2850 (a, b) | 0.3012 (c) |
| c/a (II/I) | 1.033 | | 1.042 | | 1.057 | |

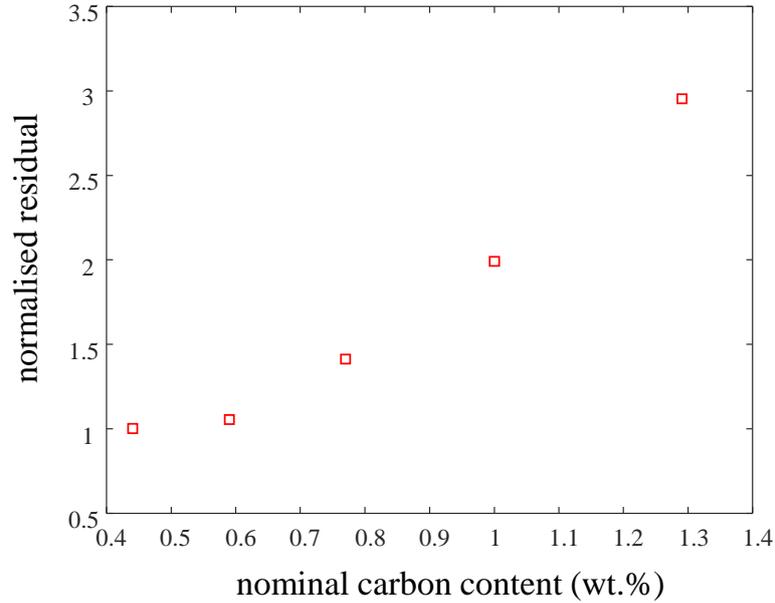

Figure 8. The fitting residual, after fitting 2 peak pairs to the experimental peak profile, with respect to nominal carbon content normalised by the residual for Fe-0.44 wt.%C martensite. High normalised residual indicates greater volume fraction of martensite with reduced tetragonality than indicated by splitting of two peak pairs.

The calculated *a, b, c*-axes lengths and the tetragonality, $(c/a)_{II/I}$ are plotted against the nominal carbon content in Fig. 9. It is found that $(c/a)_{II/I}$ well matches the *c/a* expected from eq.(1) when [C] ≥ 0.44 wt.%, showing a linear relation with respect to the nominal carbon content. This means that the classical model shown in *eq.* (1) is derived from the nearly maximum *c/a* in an XRD analysis volume while the 'averaged' *c/a* in the analysis volume should have a smaller slope, which agrees with the Rietvelt refinement analysis results [15]. On the contrary, a clear sign of tetragonality for Fe-C martensite with [C] ≤ 0.24 wt.% is not observed in this study while the Rietvelt analysis method detected the *c/a* of ~ 1.004 for 0.12 [C] martensite [15]. The extent of auto-tempering might be different between the samples used in this study and in [15]. Further comparison of our results to the Rietvelt refinement by Lu *et al* [15] is not possible as the details in the sample condition such as RT aging time and in the parameters used for the Rietvelt refinement are not mentioned. Cheng *et al* reported that the area of Peak I, $A_I$, remained unchanged while a decrease in $A_{II}$ and an increase in $A_{III}$ were observed during RT aging [27]. Therefore 'averaged' *c/a* ratio is expected to be dependent on the aging time in addition to the cooling rate.

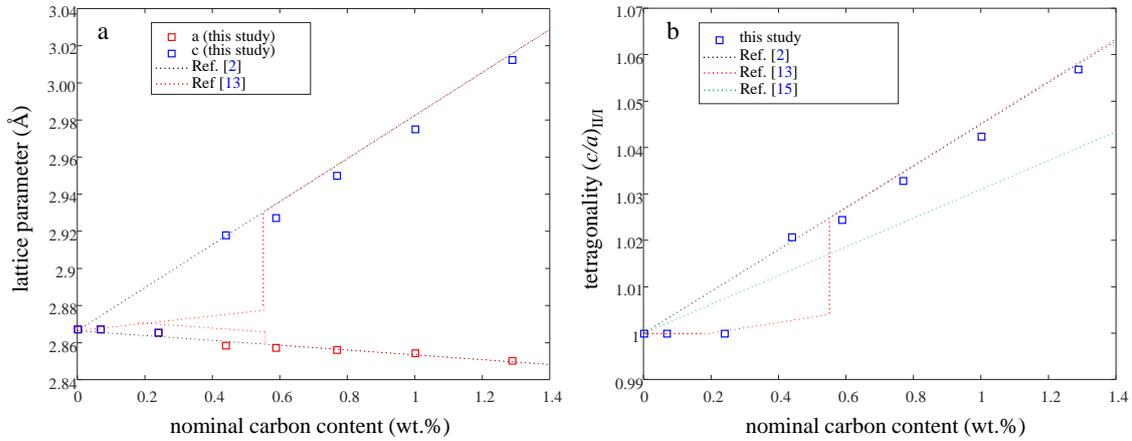

Figure 9. (a) Lattice parameters and (b) tetragonality, of Fe-C quenched and RT aged martensite determined by the $\alpha$-Fe {200} XRD Peak I&II positions.

3-2. EBSD study on the tetragonality of Fe-C quenched and RT aged martensite

Fig. 10 shows the negative of cross correlation coefficient, -$r$, as a function of imposed $c/a$ for each of three principal axes being the $c$-axis. For EBSPs from the annealed IF steel with expected $c/a$ is exactly 1, all three profiles show the minimum value of $-r$ at imposed $c/a = 1$ (Fig. 10(a)). The optimised value of $r \approx 0.73$ at $c/a = 1$ shows a similar degree of fit in each case. A similar situation was found for the other EBSP obtained from the IF steel.

In the case of 0.77 wt.% C steel, the value of $r$ is decreased from the one in Fig. 10(a), which is ascribed to the blurring of the experimental EBSPs from the martensite. Only one of the three profiles shows the minimum of $-r$ at imposed $c/a = 1.021$, meaning that the $c$-axis (axis 1 in Fig. 10(b)) and the local $c/a$ can be determined simultaneously from a single EBSP. It is noted that the other two profiles (axes 2 and 3) show the minimum of $-r$ at a similar higher level and the same imposed $c/a$ which notably is less than 1. Interpolating the $-r$ variation with $c/a$ by fitting a second order polynomial allows the minimum position to be located but also allows an estimate of the error. The error in $c/a$ can be estimated from the range of $c/a$ values for which the trend line increases from the minimum value by less than the root mean square deviation of $-r$ data from it. This amounts to ±0.002 for the IF steel data in Fig 10(a), and slightly higher at ±0.004 for the 0.77wt% C steel shown in Fig 10(b). Similar estimates were made for the other samples and show some increase in error as the C content and $c/a$ ratio increase presumably due to the poorer pattern quality that tends to be obtained.

  A second example for an EBSP obtained from the same 0.77 wt.% C steel but from different analysis point is shown in Fig. 10(c). The calculated *c/a* is different from the first example (Fig 10(b)) by ~1% indicating that the *c/a* ratio varies from analysis point to point. Again, for *c*-axis aligned along axis 1 gives the best optimisation with a *c/a* ratio above unity, while the other two cases have minima at *c/a* less than 1 but are less well optimised. For this second example all three profiles are distinct and there is less difference in the minimum values of –*r* between the three cases, which notably is not as well optimised than the example in Fig. 10(b) and corresponds to a lower *c/a* ratio. This suggests that the local crystal structure in this second case might not be best represented by 'pure' BCT as the length of all principal axes being different each other. The crystal structure can be regarded as orthorhombic or less symmetric structure. The distorted nature of the crystal structure of Fe-C quenched martensite will be further investigated by HR-EBSD strain analysis in a subsequent publication. In Fig. 10(c), the tetragonality with the minimum of –*r* among the three profiles is used to assess the local tetragonality.

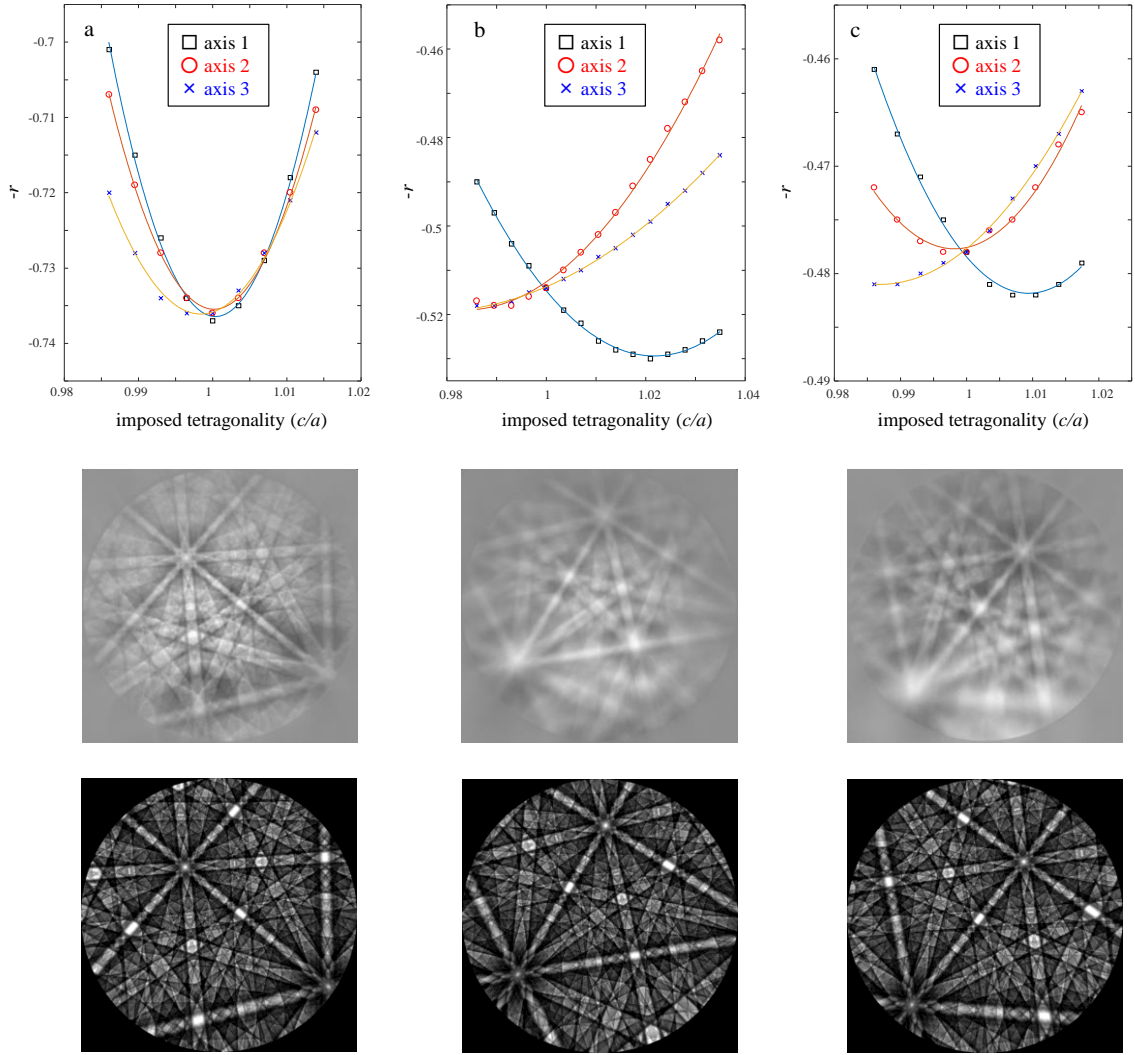

Figure 10. Negative of normalised cross correlation coefficient, -*r*, between an experimental EBSP and dynamically simulated EBSPs for BCT Fe with imposed varying tetragonality. The solid lines in the graphs show the second order polynomial fitting results. Experimental pattern is obtained from (a) IF steel, (b) Fe-0.77 wt.% C, (c) Fe-0.77 wt.% C but different analysis point from (b). Below each plot the experimental EBSP and the best fit simulation.

Fig. 11(a) summarises the calculated local *c/a* ratio as a function of nominal carbon content in Fe-C quenched and RT aged martensite. The result for the IF steel is also depicted. The averaged *c/a* for each Fe-C martensite shows that the tetragonality starts to appear from 0.44 wt.% C and a linear relation of averaged *c/a* with respect to the nominal carbon content is obtained, which is also in good agreement with the XRD results. The averaged *c/a* values found by pattern matching of EBSP are somewhat below the *c/a* expected from Eq. (1). The following is obtained for the 'averaged' tetragonality as a function of nominal carbon content with [C] ≥ 0.44 wt.%.

$$(c/a)_{ave} = 1 + 0.030 \, [C] \tag{5}$$

Again it is noted that the above relation is obtained for RT aged martensite. It should also be noted that the rejection from the analysis of EBSPs showing pattern overlap is likely to have slightly emphasised block interiors compared to regions very close to block boundaries. This probably also means that larger blocks are weighted somewhat more than smaller blocks. Similarly, the avoidance of highly blurred EBSPs means that regions with high dislocation contents and/or large strain gradients are under-represented in the set of patterns analysed. This may be related to the discrepancy between $c/a$ values reported on-average by the EBSD analysis compared to the X-ray measurements based on the positions of peaks I&II. Further analysis of the effect of boundaries and the size of blocks on the $c/a$ ratio distribution will be presented in the forthcoming publication.

The slope of 0.030 is nearly equal to the one obtained by the Rietvelt refinement analysis of XRD profile (0.031) [15]. This is probably because the Rietvelt refinement analysis optimises tetragonality so as to reproduce the whole profile of XRD pattern rather than just analysing peak positions.

The scatter in individual $c/a$ shown in Fig. 11(b) is found to be significant. The deviation from its mean value is dependent on the carbon content. This corresponds well to Fig. 8 which shows that for the XRD profiles the area of intermediate region between Peaks I and II is increased with respect to carbon content. For 1.29 wt.% C martensite, the local $c/a$ ratio ranges from 1.02 to 1.07. It is noted that the $c/a$ deviation is observed between different blocks. The spatial variation in the tetragonality within one block will be shown in a subsequent paper, incorporating cross-correlation based HR-EBSD analysis.

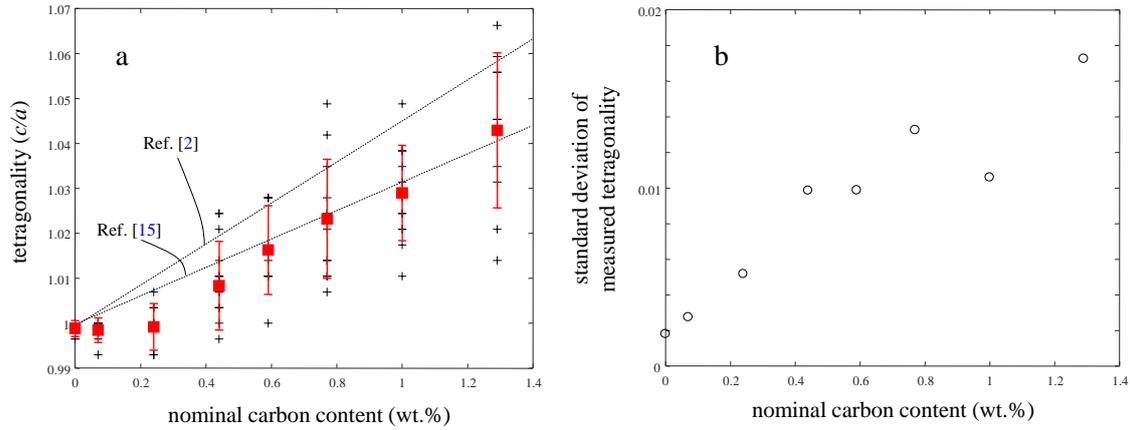

Figure 11. (a) Tetragonality determined by the pattern matching of EBSPs. The mark '+' represents the measured tetragonality at each analysis point. A red square (■) is the averaged tetragonality for each carbon content and the error bar shows ±1 standard deviation. (b) The standard deviation of measured tetragonality as a function of nominal carbon content.

One might suspect that the Ar ion sputtering during the sample preparation influences the measured *c/a* and its scatter by the ion beam bombardment-induced displacement of interstitial carbon atoms. Fig. 12 shows the local *c/a* measurement for the 1.29 wt.% C martensite prepared without Ar ion sputtering but electropolishing. In this case the PC position cannot be determined by the pattern matching of EBSPs from unstrained region as strain-free calibrant was not mounted on the specimen. The PC position determined for the previous setting by the pattern matching was just used. A two sample *t*-test was performed on the two tetragonality data sets and the mean value of the tetragonality for both cases are found to be the same to the 5 % significance level. Therefore the Ar ion sputtering condition used in this study does not significantly influence the local tetragonality measurement result.

Then there are two possible explanations for the *c/a* scatter, namely heterogeneous residual strain and carbon distribution. The carbon distribution heterogeneity can be caused by auto-tempering [12, 33, 34] and possibly by heterogeneous residual strain itself [14, 31, 32]. The scatter in the local residual strain would be more significant in high carbon martensite as it is known that Fe-C martensite becomes harder with respect to the carbon content [1] thereby allowing for taking up more elastic strain inside the material. However, the residual strain is not likely to be the sole reason for the *c/a* scatter because the scatter in *c/a* of ~ 5 % is too big to be supported. The heterogeneous carbon distribution and the consequent variation in the natural stress free lattice parameter should play some role in the *c/a* scatter.

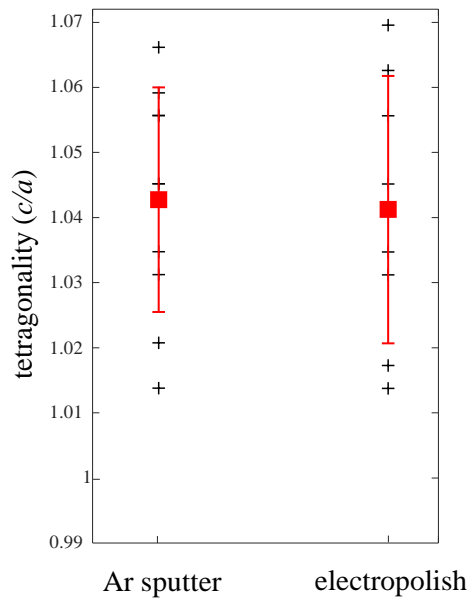

Figure 12. Tetragonality determined by the pattern matching of EBSPs using Ar sputtered specimen or electropolished specimen (specimen: Fe-1.29 wt.% C). A red square (■) is the averaged tetragonality for each carbon content and the error bar shows ±1 standard deviation.

In this study the lattice parameter for each axis is obtained by the XRD analysis while only axial ratio is characterised by the EBSD measurement. The change in the lattice parameter can be seen in EBSPs as a change in the Kikuchi band width. However the Kikuchi band edge from Fe-C quenched martensite becomes blurred because of the perturbation in the diffracting planes in an EBSD analysis volume [61]. This blurring is itself evidence of very fine-scale variations in the lattice driven by either local chemistry or mechanical stress. However, the blurred EBSP did prohibit the accurate determination of each lattice parameter by EBSD analysis which was not pursued in this study. Nevertheless, the proposed analysis approach with careful determination of PC position using a strain-free calibrant allows for the quantitative measurement of local *c/a* ratio. The tetragonality measurement is increasingly getting attention for carbon super-saturated bainite and pearlite [62-64]. The local tetragonality analysis with EBSD will shed light on the behaviour of interstitial atoms at microstructure scale during phase transformation, heat treatment and mechanical loading that need to be highly controlled for the development of advanced high strength steel.

## 4. Conclusion

In this study we explored a route to determining the local tetragonality of Fe-C quenched and RT aged martensite through the EBSD pattern matching approach incorporating careful calibration of the EBSD system. Central to the methodology is the use of a small strain-free calibrant excised by FIB from a well-annealed IF steel sample and mounted onto the surface of the strained material. This enables accurate calibration of the EBSD detector through pattern matching of EBSPs. XRD measurements were also performed to validate the EBSD analysis results. The following four findings are in good agreement between tetragonality measured by EBSD and XRD:

(1) Both XRD and EBSD tetragonality analysis show that the tetragonality starts to appear in quenched and RT aged Fe-C martensite prepared in this study as the carbon content is increased to 0.44 wt.% and beyond.
(2) Above 0.44 wt% carbon, the tetragonality determined by the XRD α-Fe {200} Peak I&II positions is linearly dependent on the carbon content while 'averaged' tetragonality determined by EBSD analysis is also increased linearly with respect to nominal carbon content.
(3) The EBSD pattern matching measurements for $c/a$ are somewhat lower than for the XRD data based on peak I&II positions. Rejecting very blurred or overlapping EBSPs from the analysis may be a factor here, although it is not proven that this necessarily biases the EBSD results to lower $c/a$ values. The XRD peak profiles showed additional intensity between peaks I&II that was not accounted for in the fitting, and increased with increasing C content. This suggests that the $c/a$ value implied by XRD peak positions may over-estimate the true volume averaged value and that the martensite is not uniformly tetragonal.
(4) The standard deviation of measured tetragonality from EBSD pattern matching increased with respect to nominal carbon content. This also corresponds to the fact that the area of peak fitting residuals is increased as the nominal carbon content increases.

## Acknowledgements

One of the authors (T. Tanaka) expresses sincere thanks to Mr. Daisuke Maeda and Dr. Yuri Kitajima (Nippon Steel Corporation) for the provision of the materials used in this study. Mr. Michiaki Matsumoto (Mishima Kosan Corporation) is also acknowledged for his help with the SEM sample preparation.